\documentclass[aps,groupedaddress,%
amsmath,preprintnumbers,10pt,prl,twocolumn]{revtex4}

\usepackage{graphicx}
\usepackage{amsfonts}
\usepackage{amssymb}
\usepackage{amsbsy}
\usepackage{amsmath}

\renewcommand*{\pl}{\text{pl}}
\newcommand*{\M}{M_\star}

\newcommand*{\eff}{\text{eff}}

\begin{document}
\preprint{ }
\title{Non-singular inflationary universe from polymer matter}
\author{Golam Mortuza Hossain}
\email{ghossain@unb.ca}
\author{Viqar Husain}
\email{vhusain@unb.ca}
\author{Sanjeev S. Seahra}
\email{sseahra@unb.ca }

\affiliation{ Department of Mathematics and Statistics, University
of New Brunswick, Fredericton, NB, Canada E3B 5A3} 
\date{October 19, 2009}

\begin{abstract}

We consider a polymer quantization of a free massless scalar field in a
homogeneous and isotropic cosmological spacetime. This
quantization method assumes that field translations are fundamentally discrete,
and is related to but distinct from that used in loop
quantum gravity. The semi-classical Friedman equation yields a universe that is
non-singular and non-bouncing, without quantum gravity. The model has an
early de Sitter-like inflationary phase with sufficient expansion to
resolve the horizon and entropy problems, and a built in mechanism for a
graceful exit from inflation.

\end{abstract}

\maketitle

\paragraph*{\bf Motivation}

The standard model of cosmology is in remarkable agreement with
current observations. The model relies   on the existence of a
sustained period of accelerated expansion in the early universe.
This inflationary epoch solves  the flatness, horizon, and entropy
problems, and as a bonus,  provides a mechanism for the generation
of primordial perturbations.  However a drawback is that this epoch
is put in by hand.

An important goal for fundamental theory therefore is to provide a
solid theoretical foundation for inflation. This has led to  much
effort especially  in string theory, although a conclusive and
widely accepted answer is not currently available. An alternative
approach known as loop quantum gravity (LQG), has also been much
studied, especially in its application  to cosmology, (known as
LQC) \cite{Ashtekar:2007tv,Bojowald:2008zzb}. One of the main
results in this  area is that the big bang singularity of general
relativity (GR) is not present in this model \cite{Bojowald:2001xe}.
In  LQC there is a a quantum bounce \cite{Date:2004fj} that occurs
when the density of matter reaches a critical value $\rho_\text{c}
\sim M_\pl^4$, where $M_\pl$ is the Planck mass
\cite{Ashtekar:2006wn}. The bounce is followed by a short
inflationary phase, but the amount of accelerated expansion is not
enough to address phenomenological questions \cite{Bojowald:2002nz,
Date:2004yz}. Thus, so far  no natural mechanism for
sufficient inflation has emerged in this approach.

Both LQG and its mini-superspace realization to cosmology are
formulated using a background independent quantization procedure
(``polymer quantization'' ), where the inner product used is
independent of the manifold metric. This is an important ingredient
since the metric is a dynamical field in quantum gravity. All the
work in LQC relies on applying this quantization to the geometric
variables, but the usual Schr\"odinger quantization is used for the
matter degrees of freedom, which is typically a homogeneous
minimally coupled scalar field.

Polymer quantization may be viewed as  a separate development in its
own right, and is applicable to any classical theory whether or not
it contains gravity  \cite{AFW, Halvor, HWI,HW-qbh1,HW-qbh2}. Its
central feature is that  momentum operators are not realized
directly as in Schr\"odinger quantum mechanics because of a built in
notion of discreteness, but arise indirectly through translation
operators. (There is a ``dual'' representation where  it is the
configuration variables that are represented only in an
exponentiated form \cite{Halvor}.)

In this Letter we show that singularity resolution, sufficient
inflation, and a graceful exit  are  all natural consequences of
classical gravity  sourced by polymer quantized matter. This
approach may be viewed as the ``polymer semiclassical
approximation.'' Unlike  LQC, where singularity avoidance results
from polymer geometry, all our results follow from solely from
polymer matter coupled to classical geometry.

The basic insight from which these results follow is that polymer
quantization of matter puts an upper bound  on its kinetic energy,
which naturally bounds curvature through the Hamiltonian constraint
of GR. How this is realized in detail is described by using the
massless  scalar field coupled to the Friedmann-Robertson-Walker
(FRW) geometry,  the same model that was studied in LQC
\cite{Ashtekar:2006wn}. However, the procedure and variables we use
for quantization are quite different from those of  LQC -- the only
common feature is the use of polymer quantization.

We assume that the scale associated with this quantization is given
by a mass parameter $\M$ (which is a priori distinct from the Planck
mass $M_\pl$). We show in this quantization that the matter energy density is
bounded: $\rho \lesssim \M^4$.  We also show that the universe approaches
a de Sitter phase in the   past for generic choices of
parameters; this nonsingular inflationary universe replaces the
bounce found in some models (such as LQC). Furthermore, for  $\M^4
\ll M_\pl^4$ the curvature of the universe remains much less
than the Planck scale, implying the quantum gravity effects are
negligible and the classical treatment of geometry is justified.
There is  a special limiting case where the universe approaches
Minkowski space in the asymptotic past.

\paragraph*{\bf Polymer quantization}

The model we consider is the massless scaler field minimally coupled
to gravity. The total Hamiltonian of this system is the  constraint
$H_g + H_\phi = 0$, where $H_g$ and $H_\phi$ are the standard
Hamiltonians for the gravity and matter sectors, respectively.
 The matter Hamiltonian is
\begin{equation}\label{SFHam}
H_{\phi} = \int d^3x N\left[ \frac{1}{2\sqrt{q}} P_\phi^2 +
\frac{\sqrt{q}}{2} q^{ab} \partial_a \phi\partial_b\phi \right],
\end{equation}
where $N$ is the lapse function, $q_{ab}$ is the spatial metric,
$q=\det(q_{ab})$, and the canonical phase space variables are
$(\phi,P_\phi)$. Instead of $(\phi,P_\phi)$, we use the basic
variables
\begin{equation}\label{BasicVariables}
\phi_f \equiv \int d^3x \sqrt{q} \, f(\textbf{x}) \phi(\textbf{x}) ,
\quad U_{\lambda} \equiv \exp\left( \frac{i \lambda
P_{\phi}}{\sqrt{q}} \right),
\end{equation}
where the smearing function $f(\textbf{x})$ is a scalar.  Such
variables are a typical choice for polymer quantization
\cite{AFW,HWI}  Since $P_\phi$ transforms as a density under
coordinate transformations, the $\sqrt{q}$ factor in the second
definition is required to make the argument of the exponent a
scalar;  {\it this factor turns out to be crucial for obtaining our results.} The parameter $\lambda$ 
is a spacetime constant with dimensions of $(\text{mass})^{-2}$.  These  variables satisfy the
Poisson algebra
\begin{equation}
\label{basic-pb} \{\phi_f,U_\lambda\} = if\lambda U_\lambda.
\end{equation}

From here on we specialize to an FRW spacetime with spatial metric
$q_{ab} = a^2(t) \delta_{ab}$ and choose the proper time gauge $N =
1$. Homogeneity requires that the smearing function be constant, so
we select $f(\textbf{x}) = 1$.  Employing a standard box
normalization to regulate the spatial integration in
(\ref{BasicVariables}) gives  the symmetry reduced variables
\begin{equation}\label{ReducedVariables}
\phi_f = V_0 a^3 \phi , \quad U_\lambda= \exp\left(i\lambda
P_{\phi}/a^3 \right),
\end{equation}
where $V_0 = \int d^3 x$ is a fiducial comoving volume. The Poisson
bracket of the reduced variables is the  same as that of the
unreduced ones (\ref{basic-pb}).

Quantization proceeds by realizing the Poisson algebra
(\ref{basic-pb}) as a commutator algebra on a suitable Hilbert
space;  the choice for polymer quantization has the basis
$\{|\lambda\rangle | \lambda \in \mathbb{R}\}$ with inner product
\begin{equation}\label{InnerProduct}
\langle\lambda^{'}|\lambda\rangle = \delta_{\lambda,\lambda^{'}},
\end{equation}
where $\delta$ is the generalization of the Kronecker delta to the
real numbers \cite{AFW}. The operators $\hat\phi_f$ and $\hat
U_\lambda$ have the action
\begin{equation}\label{operator actions}
    \hat\phi_f |\lambda\rangle = \lambda|\lambda\rangle,
    \quad \hat U_{\lambda'} |\lambda \rangle = |\lambda + \lambda'\rangle;
\end{equation}
i.e., $|\lambda\rangle$ is an eigenstate of the smeared field
operator $\hat\phi_f$, and $\hat U_\lambda$ is the generator of
field translation.

With this realization it is evident that configuration eigenstates
are normalizable. This is one of the main difference between the
polymer and Schr\"odinger quantization schemes. It is because of
this that the momentum operator does not exist in this quantization,
but must be defined indirectly using the translation generators by
the relation
\begin{equation}\label{SFMomentum}
P_{\phi}^\lambda= \frac{a^3}{2i\lambda} (U_{\lambda } -
U^{\dagger}_{\lambda }) ~.
\end{equation}

At this stage it is convenient to fix the polymer quantization scale
by setting $\lambda = \lambda_{\star} = 1/M_{\star}^2$ in the
momentum operator; we set $P_\phi^* \equiv
P_\phi^{\lambda_{\star}}$. The limit $\M \rightarrow \infty$ gives
$P_\phi^\star \rightarrow P_\phi$ at the classical level, but in
polymer quantization  $\M$ remains a fixed and finite scale.

Our prescription for cosmological dynamics is through the effective
Hamiltonian constraint
\begin{equation}\label{SCHam}
    H_g + \langle \psi | \hat{H}_\phi | \psi \rangle =0,
\end{equation}
where  $|\psi\rangle$ is a suitably chosen matter semi-classical
state, and  $H_\phi = V_0 (P_\phi^*)^2/2a^3$  in an FRW background.
Using (\ref{SFMomentum}) gives
\begin{equation}\label{rho eff defn}
    \langle \hat{H}_\phi \rangle \equiv V_0 a^3 \rho_\eff, \quad \rho_\eff  =
    \tfrac{1}{8} \M^4 [ 2 -
    \langle \hat{U}_{2\lambda_\star} \rangle - \langle \hat{U}^\dagger_{2\lambda_\star}
    \rangle ],
\end{equation}
where $\rho_\eff$ is the quantum corrected matter density. To
compute the expectation values of the translation operator, we take
$|\psi\rangle$ to be a Gaussian coherent state peaked at the phase
space values $(\phi_0,P_\phi)$. Such a state is
\begin{equation}
    |\psi\rangle = \frac{1}{\mathcal{N}} \sum_{k=-\infty}^{\infty}   c_k
    |\lambda_k\rangle, \quad
    c_k \equiv e^{-(\phi_k-\phi_0)^2/2\sigma^2} e^{-iP_\phi \phi_k V_0},
\end{equation}
where $\phi_k = \lambda_k/V_0 a^3$ is an eigenvalue of the scalar
field operator derived from $\hat{\phi}_f$ in Eq.~(\ref{operator
actions}). The scalar configuration points $\lambda_k$ are chosen
such that the Gaussian profile is well sampled; the simplest example
is a uniform sampling.

The normalization factor is  $\mathcal{N}^2 = \sum_k |c_k|^2 \simeq
V_0a^3\sigma \sqrt{\pi}$, where the symbol $\simeq$ means that the
sum is approximated by an integral.
 This state gives the expectation value
\begin{gather}\label{U EV}
    \langle \hat{U}_{\lambda_\star} \rangle \simeq
    e^{i\Theta} e^{-\Theta^2/4\Sigma^2}, \\
    \Theta \equiv \lambda_\star P_\phi a^{-3} = P_\phi \M^{-2} a^{-3}, \quad \Sigma \equiv \sigma
    V_0 P_\phi.
\end{gather}
Using this result the quantum corrected effective energy density
(\ref{rho eff defn}) is
\begin{equation}\label{EffDensity}
\rho_{\eff} (a,P_\phi;\sigma,M_\star)\simeq \tfrac{1}{4} \M^4 [1 -
e^{-\Theta^2/\Sigma^2} \cos2\Theta   ].
\end{equation}
Its dependence on  $M_\star$ is through the momentum operator, and
on $\sigma$ and $P_\phi$ through the choice of state.

The variables  $\Theta$ and $\Sigma$, and hence $\rho_\eff$, are
invariant under re-scalings of the spatial coordinates, or
equivalently the re-definition of the scale factor $a$:
\begin{equation}
    \mathbf{x} \rightarrow \ell \mathbf{x}, \quad a \rightarrow
    \ell^{-1} a, \quad V_0 \rightarrow \ell^3 V_0, \quad P_\phi
    \rightarrow \ell^{-3} P_\phi.
\end{equation}
This physically necessary property of the quantum corrected energy
density follows directly from the $\sqrt{q}$ factor in the
definition (\ref{BasicVariables}) of $U_\lambda$.  Furthermore,
$\Theta$ and $\Sigma$ have natural interpretations:
\begin{equation}
    \Theta = \sqrt{2\rho_\text{cl}/\M^4},
\end{equation}
where $\rho_\text{cl}=P_\phi^2/2a^6$ is the classical density of a
free scalar field in an FRW universe, and
 \begin{equation}
    \frac{\Delta P_\phi^2}{\langle P^\star_\phi \rangle^2} = \frac{1}{2\Sigma^2} +
    \mathcal{O}\left(\frac{\Theta^2}{\Sigma^2}\right),
\end{equation}
in the late time ($a \rightarrow \infty$) limit,  where $\Delta
P^2_\phi = \langle P_\phi^{\star 2} \rangle - \langle P_\phi^\star
\rangle^2$ is computed using (\ref{SFMomentum}) and (\ref{U EV}).
From this it is evident that as $\Delta P_\phi \rightarrow 0$,
$\Sigma \rightarrow \infty$ for fixed $\langle P^\star_\phi
\rangle$. Thus  $\Sigma$ measures how ``squeezed'' the state
$|\psi\rangle$ is in the $P_\phi$ direction at late times.

\paragraph*{\bf Effective cosmological dynamics}
Having derived an expression for the effective density
(\ref{EffDensity}), the Friedmann equation, and the evolution
equations for $\phi$  and  $P_\phi$ follow from (\ref{SCHam}). To
derive the latter, the peaking values in the state $|\psi\rangle$
are treated as a canonically conjugate pair. We find
\begin{gather}\label{eq:Friedmann}
    H^2 = \frac{\dot{a}^2}{a^2} = \frac{8 \pi G}{3}\, \rho_\eff(a,P_\phi;\sigma,M_\star),
\\
\dot{P}_\phi=0, \ \ \ \  \dot{\phi} = \tfrac{1}{2}\ \M^2
e^{-\Theta^2/\Sigma^2}\sin(2\Theta).
\end{gather}
The form of $\rho_\eff$ implies that the Hubble parameter is bounded
from above $H^2 \le 4\pi G \M^4/ 3$.  Hence, the polymer
quantization effects ensure that there is no curvature singularity
in this model.

The  behaviour of $a(t)$ may be seen from the asymptotics  of
$\rho_\eff$.  When the scale factor is large (or $\Theta \lesssim
1$) we find
\begin{equation}
    \rho_\eff \sim \frac{1}{2a^6}\left(  P_\phi^2 + \frac{1}{2\sigma^2 V_0^2}\right),
    \quad a \gtrsim a_\text{GR} \equiv \left(\frac{P_\phi}{\M^2}\right)^{1/3}.
\end{equation}
This is coincident with the limit in which polymer quantization
reduces to the Schr\"odinger one; i.e., $\M \rightarrow \infty$. In
this regime we have $H \propto a^{-3}$ and $a \propto t^{1/3}$,
which is the standard result for a massless scalar field coupled to
the FRW background in general relativity; hence, $a_\text{GR}$
indicates the beginning of an epoch where the model evolves
classically.

Conversely, when the scale factor is small (or $\Theta \gtrsim
\Sigma$) we obtain
\begin{equation}
    \rho_\eff \sim \tfrac{1}{4}\M^4, \quad a \lesssim a_\text{dS}
    \equiv \Sigma^{-1/3} a_\text{GR}.
\end{equation}
That is, the Hubble parameter is constant in the early universe,
which gives the asymptotic solution
\begin{equation}
    a(t) \propto \exp(H_\star t), \quad
    H_\star \equiv (2\pi G \M^4 / 3)^{1/2}.
\end{equation}
This is one of the main results of this work: polymer quantization
of a free scalar field coupled to an FRW universe gives an effective
de Sitter inflationary phase in the early universe that ends when $a
\sim a_\text{dS}$.  During this phase, the Hubble parameter is
incredibly close to constant; i.e., when $a = e^{-N} a_\text{dS}$ we
find $|H^2/H_\star^2 - 1| \sim \exp(-\exp 6N)$.  This exponential
expansion persists into the   past ($t \rightarrow -\infty$), so
the number of $e$-folds  is   infinite.

\begin{figure*}
\begin{center}
    \includegraphics[width=\textwidth]{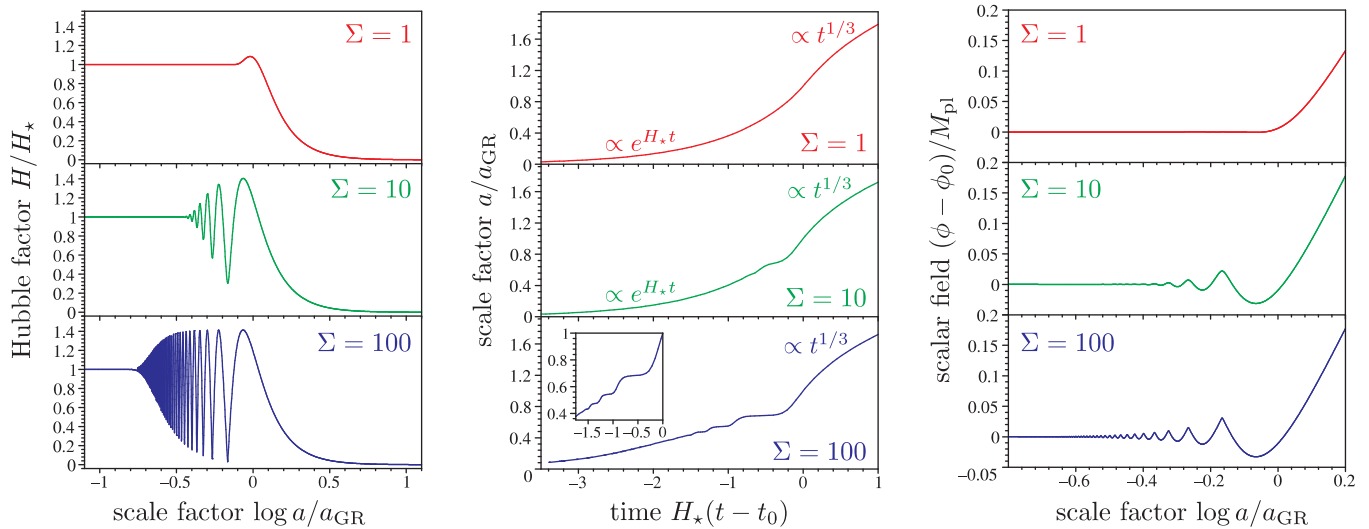}
\end{center}
\caption{Behaviour of the Hubble parameter (\emph{left}), scale
factor (\emph{center}) and field amplitude (\emph{right}) for
various values of $\Sigma$.  Here, $\phi_0$ and $t_0$ are arbitrary
constants. We see that $H$ is virtually constant for $a \lesssim
a_\text{dS} = \Sigma^{-1/3} a_\text{GR}$, which leads to an early
time de Sitter phase where $a$ grows exponentially and $\phi$ is
constant. Between $a_\text{dS}$ and $a_\text{GR}$, $H$ and $\phi$
show oscillatory behaviour while $a$ evolves in a ``stair-step''
pattern.  When $a \gtrsim a_\text{GR}$, we recover conventional
classical dynamics.}\label{figure}
\end{figure*}

To summarize:  The universe undergoes a de Sitter-like expansion
when $a \lesssim a_\text{dS}$, or when the classical energy density
is large $\rho_\text{cl} \gtrsim \Sigma^2 \M^4$.  When $a \gtrsim
a_\text{GR}$, or $\rho_\text{cl} \lesssim \M^4$, the universe
evolves as in GR. In Fig.~\ref{figure}, we illustrate these features
of the cosmological dynamics for several specific choices of
$\Sigma$.

\paragraph*{\bf Emergent universe}

The effective early time de Sitter phase  is  
present when the squeezing parameter $\Sigma$ is finite, but otherwise independent
of its actual value.   If $\Sigma
\rightarrow \infty$, the Friedmann equation reduces to $H^2 = ({4\pi
G \M^4}/{3}) \sin^2 \Theta$.  This has a number of \emph{static}
solutions:
\begin{equation}
    a(t) = \left({P_\phi}/{\M^2 n \pi}\right)^{1/3} \equiv a_n, \quad \dot{a}(t)
    = \ddot{a}(t) = 0,
\end{equation}
where $n = 1,2,3\ldots$  Each of these solutions are unstable fixed
points representing \emph{Minkowski} 4-geometries. In particular,
one can find scale factor solutions with asymptotic behaviour
\begin{equation}\label{emergent solutions}
    a(t) \sim \begin{cases} c_1 e^{
    \sqrt{3 \pi^3 G} \M^2 t} + a_1, & t \rightarrow -\infty \\
    c_2 t^{1/3}, & t \rightarrow +\infty \end{cases},
\end{equation}
where $c_1$ and $c_2$ are constants.  This class of solutions
reproduces the conventional universe dynamics at late times;
however, at early times the universe asymptotes to Minkowski space
with $a \rightarrow a_1$.   We comment that the solutions
(\ref{emergent solutions}) are reminiscent of the emergent universe
scenario proposed by \citet{Ellis:2002we} where the initial state
was an Einstein-static configuration with positive spatial
curvature.

\paragraph*{\bf Discussion}

We have explored the cosmological consequences of the idea that
quantum translations of the amplitude of a free scalar field are
fundamentally discrete.  Working with semi-classical quantum states,
we found that this simple principle leads to the avoidance of the
big bang singularity when such fields are coupled to a homogeneous
and isotropic classical universe. Unlike LQC which uses a similar
quantization scheme for gravitational degrees of freedom, universes
containing polymerized scalar fields approach de Sitter or Minkowski
space in the asymptotic past. Thus  inflationary or flat universes
are past attractors of this model, with classical general relativity
recovered at late times.

As can be seen in the effective Friedmann equation
(\ref{eq:Friedmann}), our model is characterized by two mass scales.
The first of the these $\M$ is a fundamental parameter of our
quantization procedure, and is directly related the magnitude of
discrete field translations.  The second parameter $\sigma$ gives
the quantum uncertainty in the field amplitude for the
semi-classical states we are considering.  As long as $\sigma$ is
finite, we are guaranteed that the universe will undergo inflation
at early times. If we assume GUT-scale inflation with $\M \sim
10^{15} \, \text{GeV}$, we are guaranteed that the maximum density
achieved by the scalar field is much less than $M_\pl^4$, and we may
safely neglect the quantum gravity effects predicted by loop quantum
cosmology.

Our model has a number of cosmologically attractive features:  In
contrast to LQC, the de Sitter-like phase is past eternal, so there is 
sufficient inflation  to solve the horizon and entropy
problems.  (However, this may imply our spacetime is past incomplete
\cite{Borde:2001nh}, which requires further study.)  Unlike many
other models, there is a natural end to the inflationary period when
the polymer quantization effects become sub-dominant. We do not have
to fine-tune parameters to obtain inflation; i.e., just assuming a
finite width for the matter quantum state is enough to guarantee the
existence of the de Sitter-like phase. Furthermore there  exists 
an interesting transition epoch between the end of the inflationary phase and
the onset of the classical period where the scale factor evolves in a pseudo-discrete
manner and the scalar field oscillates. The latter may be relevant
for reheating the universe via parametric resonance.

An open question in this scenario is the generation of primordial
perturbations.  The fact that we have nearly de Sitter
inflation would suggest that the spectrum of fluctuations produced
by this model would compare favourably with observations of the
cosmic microwave background, etc.  However, some caution is
warranted: The polymer quantization procedure will have a
significant impact on the behaviour of inhomogeneous $\phi$
perturbations.  An initial study of the dynamics of an inhomogeneous
scalar field in the context of polymer quantization can be found in
Ref.~\cite{Hossain:2009vd}, where the behaviour of $\phi$ in a
Minkowski background is studied.  We find that conventional flat
space Klein-Gordon equation is replaced by a nonlinear wave
equation, and polymer corrections to conventional dynamics depend on
both that amplitude and frequency of matter waves.  We expect this
feature to generalize to the cosmological case. It is worthwhile
mentioning that one can derive a lower bound of $\M \gtrsim
1\,\text{TeV}$ from applying the flat space results in
\cite{Hossain:2009vd} to the proton-antiproton beam in the Large
Hadron Collider.

We are supported by NSERC of Canada and AARMS.


\bibliographystyle{apsrev}
\bibliography{polymer-cosmology}

\end{document}